\documentclass[aps,prb,twocolumn]{revtex4}
\newcommand{\Sch}{Schr\"odinger}
\newcommand{\ie}{{\it i.e.}\ }

\usepackage{amsmath}
\usepackage{graphicx}
\newcommand{\ba}{\mathbf{a}}
\newcommand{\tr}{\mathop{\rm tr}}
\newcommand{\Tr}{\mathop{\rm Tr}}
\newcommand{\re}{\mathop{\rm Re}}
\newcommand{\im}{\mathop{\rm Im}}
\begin{document}
\title{Scattering Approach to Counting Statistics in Quantum Pumps}
\author{B.A.~Muzykantskii}
\email{boris@warwick.ac.uk}
\affiliation{University of Warwick, Coventry England}
\author{Y.~Adamov}
\email{adamov@tkm.physik.uni-karlsruhe.de}
\affiliation{Institut f\"ur Nanotechnologie, Forschungszentrum Karlsruhe, Germany}

\begin{abstract}
  
  We consider the Fermi gas in a non-equilibrium state obtained by
  applying an arbitrary time-dependent potential to the Fermi gas in the
  ground state. We present a general method that gives the quantum
  statistics of any single-particle quantity, such as the charge, total
  energy or momentum, in this non-equilibrium state.  We show that the
  quantum statistics may be found from the solution of a matrix
  Riemann-Hilbert problem. We use the method to study how the finite
  measuring time modifies the distribution of the charge transferred
  through a biased quantum point contact.
\end{abstract}

\maketitle

\section{Introduction}
From a theoretical point of view, a quantum pump\cite{pumps-experiment}
at zero temperature simply excites the Fermi gas from the ground state
into a non-equilibrium state.  The exact many particle state obtained
after a few pumping cycles depends on the applied time-dependent
potential.  When the external potential changes slowly (this is commonly
referred to as ``adiabaticity'' in the context of quantum pumps) the
final non-equilibrium state is completely determined by the evolution of
the scattering matrix at the Fermi energy $S(t)$.  Therefore, it should
be possible to obtain the distribution functions of various operators
(like total energy, total momentum, total charge in one of the leads,
etc) as functionals of $S(t)$. This turns out to be a non-trivial
problem, which we address in the present paper.

Historically, the first quantity studied in the above setting was the
total energy absorbed by the Fermi gas as a result of a sudden
introduction of a single $s$-scatterer. The distribution function for
absorbed energy shows a power-like divergence at low energies -- the so
called Fermi edge singularity (see e.g. Ref.~\onlinecite{Mahan-book} for
details). The distribution of the absorbed energy for a more general
situation, when the strength of the $s$-wave scatterer changes
arbitrarily in time, was considered in \cite{BrakoNewns:81,Makoshi}.  In
all these cases the scattering matrix $S(t)$ commutes with itself at
different times $ S(t) S(t') = S(t') S(t)$. The non-commutative case,
surprisingly, is much more difficult
\cite{Makoshi_S,adamov-muzykantskii,MoskaletsBut02}. The first non-perturbative treatment of the energy
distribution for this case \cite{adamov-muzykantskii} mapped the problem
into a one-dimensional chiral determinant, and seemed not to be easily
adapted for the computation of other quantities. In this paper we present
an alternative approach which is much simpler and also readily applicable
to distribution functions of {\bf any} single particle operator.

We illustrate our method using the distribution of the charge transmitted
through a contact, which has attracted a lot of attention
recently\cite{Lev02}. The second moment of the charge distribution (\ie
the shot noise power) for a biased quantum point contact was considered
theoretically by Khlus \cite{Khlus87} and Lesovik \cite{LesovikIntro89},
and the full counting statistics was obtained by Lesovik and Levitov (LL)
\cite{LevitovJETP93}. The charge counting statistics for quantum pumps
and quantum point contact under some specially chosen time-dependent bias
were also
considered~\cite{IvaLL97,Andreev-Kamenev2000,Makhlin-Mirlin2001}.

The experimental measurements of the shot noise
\cite{LiTsuishot90,ReznikovHeiblum95,Kumar96,SchoelkopfKozhevnikov98} are
now well established.  The measurement of the higher cumulants, on the
other hand, is still an unresolved problem actively discussed in the
literature\cite{LevitovReznikov02}.

The question of what can really be measured in an experiment, and the
closely related question of the back influence of a detector onto the
measured system are still widely discussed in the literature
\cite{LevitovReznikov02,LesovikLoosen97,GavishImry02}. We use
conceptually the simplest measurement scheme when the total charge $Q$ in
the left lead is measured twice. Firstly before the pumping is started at
time $t=0$, and secondly after the pumping is finished at $t=t_f$. At
both times the leads are decoupled, so $Q$ is a conserved quantity. The
first measurement projects onto the state with a fixed number of
electrons in the left lead. After that the pumping excites this state
into a non-equilibrium state that is no longer an eigenstate of the
charge operator. The statistics of the transmitted charge is given by
the probability to find a particular number of electrons
in the left lead during the second measurement.

Alternative approaches discussed in the literature include coupling the
system to a spin and measuring the spin precession angle\cite{Levitov96};
or measuring the statistics of the photon emission\cite{Beenakker2001};
or measuring total dissipated energy\cite{adamov-muzykantskii}. See also
Ref.~\onlinecite{NazarovKindermann01} for a general framework for
measuring the full counting statistics of a quantum mechanical variable.

Our main result is the closed expression for the full counting statistics
in terms of the scattering matrix $S(t)$ and the solution of an auxiliary
matrix Riemann-Hilbert~(RH) problem. We assume that $S(t)$ changes
adiabatically, but otherwise do not restrict its time dependence. To
demonstrate the power of the method we analyze in some detail the case of
a biased quantum contact.  In the limit of infinitely long pumping time
the charge distribution is binomial\cite{LevitovJETP93}. We compute the
corrections to the binomial distribution when the pumping time is long
but finite and discuss the physical meaning of these corrections.

The rest of the paper is organized as follows. In
section~\ref{sec:method} we relate the counting statistics to a
determinant of a certain operator; in section~\ref{sec:RH} we compute
this determinant using a solution of a matrix RH problem; in
section~\ref{sec:contact} we apply the general method to a biased quantum
point contact and rederive LL's answer. In section~\ref{correction} we
asymptotically solve the relevant RH problem and obtain corrections to
LL's answer due to finite measuring time. Some technical results are
relegated to the appendix.

\section{Distribution Function of Transmitted Charge}
\label{sec:method}
\subsection{Notations}
The free Hamiltonian $H_0$ describes non-interacting Fermi gas in two
uncoupled leads.  Let the single particle states $\psi_{n \epsilon}$ be
the eigenvalues of the Hamiltonian $H_0$ with energy $\epsilon$, where
$n=1 \ldots N$ labels energy degenerate states in both the left and the right
lead. In the simplest case when the leads are one dimensional $n=1$ for
the left lead and $n=2$ for the right lead.  The annihilation operator in
the state $\psi_{n\epsilon}$ is denoted by $a_{n\epsilon}$.  We assemble
$N$ such operators in the $N$-dimensional column $\ba_{\epsilon}$
\begin{equation}
  \label{eq:a_k}
  \ba_{\epsilon}=
  \begin{pmatrix}
    a_{1\epsilon}\\ \vdots \\ a_{N\epsilon}
  \end{pmatrix}
\end{equation}
and use the notation $\ba^+_\epsilon \equiv ( a^+_{1\epsilon}, \ldots, a^+_{N\epsilon})$ for
the $N$-dimensional row of the creation operators. This gives for the free
Hamiltonian
\[
H_0= \sum_\epsilon \epsilon \ba^+_\epsilon \ba_\epsilon.
\]
The time-dependent potential that couples the leads is described by an $N
\times N$ matrix $M(t,\epsilon, \epsilon')$, so the full Hamiltonian has
the form
\[
H(t) = H_0 + \sum_{\epsilon,\epsilon'} \ba^+_\epsilon M(t,\epsilon,\epsilon') \ba_{\epsilon'}.
\]
Analogously, the total charge operator $Q$ in the left lead is given by 
\begin{equation}
  \label{eq:Q}
  Q=\sum_\epsilon  \ba^+_\epsilon L \ba_\epsilon,
\end{equation}
where $L$ is the projector on the states in the left lead, and charge is
measured in units of electron's charge. For two one-dimensional leads
$L= \left(\begin{smallmatrix}
    1 & 0 \\
    0 & 0
\end{smallmatrix}\right).$ 

\subsection{The Characteristic Function}
We want to compute the probability $P(q)$ that the charge in the left
lead changes by $q$ during the time $t_f$. The characteristic function
$\chi(\lambda)$ is defined as
\begin{equation}
\label{eq:chi-0}
\chi(\lambda)= \sum_q P(q) e^{- i \lambda q}.
\end{equation}
At zero bias the many particle ground state $|0\rangle$ corresponds to
filling the states below the Fermi energy in both leads 
\begin{equation}
  \label{eq:ground}
  |0\rangle = \prod_{\epsilon<0} \prod_n a^+_{n\epsilon}|\rangle,
\end{equation}
where $|\rangle$ is the true vacuum state with no electrons.  We assume
that $\epsilon$ takes some discrete values, so the total number of
electrons is finite, and also set the Fermi energy to zero.  At zero
temperature we have
\[
\chi(\lambda) = \langle 0 | U^+ e^{-i \lambda Q} U e^{i \lambda Q}
|0\rangle.
\]
where $U = U(t_f)$ is  the many particle evolution operator which is the
solution of the \Sch\ equation
\[
i \frac{dU}{dt} = H(t) U(t), \quad U(0)=1
\]
Since the ground state~(\ref{eq:ground}) is the eigenstate of the
operator $Q$, the characteristic function $\chi$ can be written as
\begin{equation}
\label{eq:chi-1}
\chi=\langle 0 | U^+ e^{-i\lambda Q} U | 0 \rangle \langle 0| e^{i\lambda Q} |0\rangle.
\end{equation}

\subsection{One Particle Scattering Problem}
To proceed, we separate the single-particle scattering problem from the
computation of the expectation value over the ground state of the
Fermi-gas \cite{BrakoNewns:81,adamov-muzykantskii}.  Since the
Hamiltonian $H(t)$ is quadratic, we have
\begin{equation}
  \label{eq:motion}
  U \ba_\epsilon^+ |\rangle = \sum_{\epsilon'} \sigma(\epsilon,\epsilon') \ba_{\epsilon'}^+ |\rangle,
\end{equation}
where $\sigma(\epsilon,\epsilon')$ is some $N \times N$ matrix that fully
describes the {\bf single particle} time-dependent scattering problem. We
introduce the operator $\hat \sigma$ with matrix elements
$\sigma(\epsilon,\epsilon')$ that acts on the full space of all possible
states (\ie in both channel space and energy-space).  This operator is
unitary $\hat \sigma^{-1} = \hat \sigma^+$. We show that $\sigma$ can be
related to the scattering matrix $S(t,\epsilon)$ on the instant value of
the potential $M(t)$ for electron with energy $\epsilon$.

Consider the scattering states with energy $\epsilon$ propagating towards
the contact $\psi^{in}_{n\epsilon}$ and out of it
$\psi^{out}_{n\epsilon}$. Let us label them and fix their phases in such
a way that far from the contact $\psi_{n\epsilon}=\psi^{in}_{n\epsilon} +
\psi^{out}_{n\epsilon}$, where $\psi_{n\epsilon}$ is the eigenfunction
for decoupled electrodes introduced previously. In the presence of the
potential $M(t)$ the eigen space corresponding to the energy $\epsilon$
is still $N$-times degenerate. Far from the contact an eigen function has
asymptotic form $\sum_n a_n \psi^{in}_{n\epsilon} + b_n
\psi^{out}_{n\epsilon}$. The scattering matrix $S(\tau, \epsilon)$
relates the amplitudes of incoming and outgoing states $b_n = \sum_m
S(\tau,\epsilon)_{nm} a_m$. Note, that the reflection coefficients are
diagonal matrix elements of $S$ and our choice of phases for the
scattering states ensures that for decoupled leads $S=1$.  For two
one-dimensional leads
\begin{equation}
  \label{eq:S2}
  S=
  \begin{pmatrix}
    B & A \\ A & -B^* \frac{A}{A^*}
  \end{pmatrix}
\end{equation}
where $A$ is the transmission amplitude and $B$ is the reflection
amplitude for the electrons incident from the left. Eq~(\ref{eq:S2})
implies that the scattering potential is real (i.e. there is no magnetic
field) and uses this fact to relate the scattering amplitudes for the
electrons coming from the right and from the left. For example,
consider the two half-infinite one-dimensional conductors occupying left
and right half-line and separated by an infinite potential wall of width
$l$. Our choice of scattering states dictates $A=0$ and $B=1$.  When
the conductors are connected and potential wall is completely removed we
get $B=0$ and $A=e^{i l p/\hbar}$, where $p$ is the momentum of electrons
with energy $\epsilon$. 

The choice of the scattering states is not unique. Another (more common)
approach is to choose the scattering states in such a way that the
scattering matrix for the fully opened channel is one.  Denoting the
scattering matrix in this representation by $\tilde S$ we see that
$\tilde S= S S_{open}^{-1}$ where $S_{open}$ is the scattering matrix for
the fully open channel in the representation adopted in this paper.

The adiabaticity condition (common in describing ``quantum pumps'') means
that $S(\tau,E)$ changes slowly both as a function of time and energy
\begin{equation}
  \label{eq:adiabatic}
  \hbar \frac{\partial S^{-1} }{\partial t} \frac{\partial S}{\partial E} \ll 1.
\end{equation}
Using~(\ref{eq:adiabatic}) one can show (see e.g.
[\onlinecite{adamov-muzykantskii}]) that
\begin{equation}
\label{eq:sigma}
 \hat \sigma = \hat S 
\end{equation} 
when $\sigma$ acts on states near the Fermi energy.  Here the operator
$\hat S$ is the Fourier transform of the scattering matrix, \ie it has the
matrix elements
\begin{equation}
\label{eq:Shat}
\hat S_{\epsilon\epsilon'} = \frac{1}{2\pi\nu} \int dt S(t) e^{i(\epsilon-\epsilon')t},
\end{equation}
where $\nu$ is the density of states near the Fermi energy per single
channel. In the rest of the paper we often consider operators acting on
both energy and channel indexes. For example, in addition to $\hat
\sigma$ and $\hat S$ introduced above we use the operator $\hat L$ that
has matrix elements $L \delta_{\epsilon \epsilon'}$.  We use hats to
distinguish these operators from the corresponding matrix-valued
functions. The notation ``$\Tr$'' is used for trace in both channel and
energy space, while the symbol ``$\tr$'' is reserved to denote the trace
in the $N \times N$ channel space only.

\subsection{Averaging Over the Fermi Distribution}
The remaining task is to compute the average with respect to the Fermi
ground state~(\ref{eq:ground}).  We combine~(\ref{eq:motion}) with
(\ref{eq:Q}) and~(\ref{eq:chi-1}) and introduce the operator
\begin{equation}
  \label{eq:hatR}
  \hat R =  \hat \sigma^+ e^{-i \lambda \hat L} \hat \sigma = 
        e^{-i \lambda \hat  \sigma^+ \hat L \hat \sigma} 
\end{equation} 
to obtain
$$
U^+ e^{-i \lambda Q} U \ba^+_\epsilon |\rangle = \sum_{\epsilon'} R(\epsilon,\epsilon') \ba^+_{\epsilon'}
|\rangle.
$$

The Fermi distribution at zero temperature can be viewed as the diagonal
operator $\hat f$ with matrix elements $f(\epsilon,\epsilon') =
\delta_{\epsilon \epsilon'} \theta(-\epsilon)$. In the block notation
that separates the states with positive and negative energies we have
\[
\hat f= \begin{pmatrix}
  \hat 1 & 0 \\
  0 & 0
  \end{pmatrix}.
\]
In the same block notation the operator $\hat R$ has the form
\begin{equation}
  \label{eq:block1}
  \hat R = 
  \begin{pmatrix}
    \hat R_{11} & \hat R_{12} \\
    \hat R_{21} & \hat R_{22} 
  \end{pmatrix}.
\end{equation}
Since the ground state~(\ref{eq:ground}) is simply a Slater determinant
we have $\langle 0 | U^+ e^{-i\lambda Q} U | 0 \rangle = \det \hat
R_{11}$ for the first term in~(\ref{eq:chi-1}).  This expression is
further simplified by noting that the operator $\hat 1- \hat f + \hat f
\hat R$ has the same determinant as $\hat R_{11}$ because in the block
notation it takes the form
\begin{equation}
\label{eq:block2}
\begin{pmatrix}
\hat R_{11} & \hat R_{12} \\
0 & \hat 1  
\end{pmatrix}.
\end{equation}
The second term in~(\ref{eq:chi-1}) is simplified because $|0\rangle$ is
an eigenstate of the charge operator $Q$
\[ 
\langle 0| e^{i\lambda Q} |0\rangle = e^{i\lambda \langle 0| Q |0\rangle } =  e^{ i \lambda \Tr \hat L \hat f}.
\]
In this way we obtain
\begin{equation}
  \label{eq:chi-c2}
  \chi = \det (\hat 1- \hat f + \hat f \hat R ) e^{ i \lambda \Tr \hat L \hat f}
\end{equation}

Eq.~(\ref{eq:chi-c2}) combined with~(\ref{eq:hatR}) and~(\ref{eq:sigma})  gives
\begin{equation}
  \label{eq:chi-c3}
  \chi = \det (1- \hat f + \hat f \hat S^{-1} e^{-i \lambda \hat L}
  \hat S) e^{i \lambda  \Tr \hat L \hat f}.
\end{equation}
In the time representation $\hat S$ and $\hat L$ are multiplications by
the matrix-valued functions $S(t)$ and $L$ respectively and $\hat f$ has
matrix elements
\begin{equation}
  \label{eq:f-time}
  f(t,t')=\frac{i}{2\pi} \frac{1}{t-t' + i0}.
\end{equation}
The representation of counting statistics for a biased quantum point
contact through the determinant of type ~(\ref{eq:chi-c3}) was obtained
in Ref.~\onlinecite{IvaLL97} and later adopted for parametric
pumping\cite{Andreev-Kamenev2000} and re-derived in different
ways\cite{Klich02}. Our derivation is close in spirit to
Ref.~\cite{IvaLL97} and is probably the simplest. In addition it leads
to a simple regularisation procedure which we discuss below.

\subsection{Biased Quantum Pump}
Finally, we briefly mention that in the case when the left lead is biased
with respect to the right lead by a time-dependent potential V(t), a
gauge transformation can be performed on all states in the left electrode
\[
\ba_k \to e^{i L \int_0^t V(\tau) d\tau} \ba_k.
\]
In the new gauge the ground state is again given by~(\ref{eq:ground}).
The distribution function for this case can therefore be recovered
from~(\ref{eq:chi-c3}) by substituting the gauge transformed scattering
matrix
\begin{equation}
  \label{eq:S-gauged}
  e^{i L \int_0^t V(\tau) d\tau} S e^{-i L \int_0^t V(\tau)
    d\tau}
\end{equation}
instead of $S$.

\subsection{Regularisation} 
Eq.~(\ref{eq:sigma}) is only valid for the states near the Fermi surface.
Since the determinant in~(\ref{eq:chi-c2}) contains a non-vanishing
contribution from the states deep below the Fermi surface, the transition
from (\ref{eq:chi-c2}) to~(\ref{eq:chi-c3}) is not quite correct. In
particular, the determinant in~(\ref{eq:chi-c3}) is formally infinite
(and so requires regularisation) while Eq.~(\ref{eq:chi-c2}) gives a
finite answer. To find out the correct regulator we use the decomposition
\begin{align}
  \label{eq:reg0}
  \ln \chi  &= \Tr \left( \ln ( \hat 1- \hat f + \hat f \hat R) - 
\hat f \ln  \hat R \right)  \\
 & + \Tr (\hat f \ln \hat R)  +  i \lambda \Tr \hat L \hat f. \nonumber
\end{align}
In the first term in~(\ref{eq:reg0}) both the contributions from the
states deep below the Fermi surface (when $f=1$) and the states high
above the Fermi surface (where $f=0$) are zero, so Eq.~(\ref{eq:sigma})
can be used.  The two last terms in~(\ref{eq:reg0}) can be re-arranged to
give $ - i \lambda \Tr \left ( [ \hat \sigma, \hat f] \hat \sigma^+ \hat
  L \right) \equiv -i \lambda Q_{av}$ where the square brackets denote
the commutator. Again the contribution from the states away from the
Fermi energy vanishes and we obtain
\begin{equation}
  \label{eq:Q-av}
  Q_{av} = \Tr \left ( [ \hat S, \hat f] \hat S^{-1} \hat L \right)  =
  \frac{i}{2 \pi } \int \tr ( \frac{d S}{d t} S^{-1} L) dt
\end{equation}
To obtain the last equality in~(\ref{eq:Q-av}) we compute the trace in
the time representation, where all operators, except $f$ are diagonal,
and matrix elements of $f$ are given by (\ref{eq:f-time}). Comparing
~(\ref{eq:Q-av}) with Brouwer's formula \cite{brouwer98} for the average
charge transmitted through a quantum pump, we identify $Q_{av}$ with the
average value of the total transmitted charge.

Combining these results together we get
\begin{equation}
  \label{eq:chi-c4}
  \ln \chi = \Tr \left\{ \ln \left( \hat 1- \hat f + \hat f \hat R
 \right) - \hat f \ln \hat R \right\}  - i \lambda Q_{av}.
\end{equation}
where in the time representation $\hat R$ is multiplication by the matrix
valued function
\begin{equation}
  \label{eq:R}
  R(t) = S^{-1}(t) e^{- i \lambda L} S(t).
\end{equation}
The first term in Eq~(\ref{eq:chi-c4}) contains the information about the
second and higher moments of the charge distribution.

\section{Relation to  a Riemann-Hilbert problem} 
\label{sec:RH}
In this section we relate the operator trace
\begin{equation}
  \label{eq:Tr}
  Tr \equiv \Tr \left( \ln ( \hat 1- \hat f + \hat f \hat R) - \hat f \ln \hat R
\right)
\end{equation}
from Eq.~(\ref{eq:chi-c4}) to a solution of an auxiliary $N \times N$
matrix Riemann-Hilbert problem.
\subsection{General case}
We use
\begin{equation}
\label{eq:tr2}
Tr = \int_0^\lambda d\lambda \Tr \left[  \left( (1-\hat f + \hat f \hat R)^{-1} \hat f
  - \hat f \hat R^{-1} \right) \frac{d R}{d \lambda} \right].
\end{equation} 
To invert $1 - \hat f + \hat f \hat R$ we need to solve the following
auxiliary RH problem. Find a matrix-valued function $Y(t)$ of the complex
variable $t$ such that $Y$ 
\begin{itemize}
\item is analytic on the complement of the cut $(0,t_f)$
\item obeys
\begin{equation}
  \label{eq:RH-2}
 Y_-(t) Y_+^{-1}(t) =R(t) e^{i\lambda L}  \hbox{ when } t \in (0,t_f),
\end{equation}
where $Y_\pm(t)=Y(t\pm i 0)$ 
\item tends to $1$ at infinity
\[
Y  \to  1 \hbox{  when } |t| \to \infty. 
\]
\end{itemize}
Using~(\ref{eq:f-time}) and analyticity of $Y_\pm$ we observe that $Y_\pm$
has the following properties
\begin{eqnarray*}
  \label{eq:R-pm}
  && \hat f \hat Y_- \hat f = \hat f \hat Y_-   \\
  && \hat f \hat Y_+ \hat f = \hat Y_+ \hat f.  
\end{eqnarray*}

Using the above and substituting $\hat R = \hat Y_- \hat Y_+^{-1} e^{- i
  \lambda \hat L} $ one can verify
\[
  \left( 1 - \hat f + \hat f \hat R \right)^{-1} = e^{i \lambda \hat L}
  \hat Y_+ \left( (1-\hat f)  \hat Y_+^{-1} e^{-i \lambda L} 
    + \hat f \hat Y_-^{-1} \right)
\]
which after substitution in~(\ref{eq:tr2}) gives
\begin{equation}
  \label{eq:tr3}
  Tr = \int_0^\lambda \frac{d\lambda}{2\pi} \int dt \tr \left( e^{i\lambda L} 
  \frac {d Y_+}{dt} Y_+^{-1} e^{-i \lambda L} 
  S^{-1} L S \right)
\end{equation}
Combining~(\ref{eq:tr3}) with~(\ref{eq:chi-c4}) and~(\ref{eq:Q-av}) we
obtain our main result
\begin{eqnarray}
  \label{eq:chi-general}
  \ln \chi &=&  \int_0^\lambda \frac{d\lambda}{2\pi} \int dt \\
&& \tr \left\{
  \frac{d \left( S e^{i\lambda L} Y_+ \right) }{dt}  \left( S e^{i\lambda
  L} Y_+ \right)^{-1} L \right\} \nonumber
\end{eqnarray}
where $Y$ is analytic in the complement of the cut $(0,t_f)$, obeys $
Y_-Y_+^{-1} = S^{-1} e^{-i \lambda L} S e^{i\lambda L} $ along the cut
and goes to $1$ at infinity.

\subsection{Charge Transfer per Cycle for Periodic Pumping}
\label{sec:periodic}
Eq.~(\ref{eq:chi-general}) can also be used when the scattering matrix is
periodic $ S(t)= S(t+T)$ and we are interested in the limit of the large
number of cycles (\ie $t_f/T \gg 1$). The RH problem in this case can be
somewhat simplified by rolling the infinite real axis into a unit circle
with the mapping $z = e^{2\pi i t/T}$. After that the RH
problem~(\ref{eq:RH-2}) reduces to finding
two functions $Y_\pm(z)$ such that
\begin{itemize}
\item $Y_+$ is analytic for $|z|<1$ and $Y_-$ is analytic for $|z|>1$
\item on the unit circle $|z|=1$
\begin{equation}
  \label{eq:RH-circle}
   Y_- Y_+^{-1}=S^{-1}(z) e^{-i\lambda L} S(z) e^{i \lambda L}
\end{equation}
\item 
  $ Y_- \to 1 \hbox{ when } |z| \to \infty $
\end{itemize}

The distribution of charge transmitted during time $t_f$ in the limit
$t_f/T \gg 1$ is now given by
\begin{eqnarray}
  \label{eq:chi-periodic}
  \ln \chi &=&  \frac{t_f}{T} \int_0^\lambda \frac{d\lambda}{2\pi}
  \oint_{|z|=1} dz   \\
&& \tr \left\{  \frac{d \left( S e^{i\lambda L} Y_+ \right) }{dz}  
 \left( S e^{i\lambda  L} Y_+ \right)^{-1} L \right\}  \nonumber
\end{eqnarray}

\section{Counting Statistics for a Biased Quantum Point Contact}
\label{sec:contact}

In this section we apply the technique introduced above to the quantum
point contact described by the scattering matrix~(\ref{eq:S2}) and biased
by the time-independent potential $V$.

\subsection{General formulas}

After the gauge transformation~(\ref{eq:S-gauged}) the scattering matrix
takes the form 
\begin{equation}
\label{eq:S3}
S(t) = e^{i L V t} S_0 e^{-i L V t},
\end{equation} 
where the time-independent matrix~(\ref{eq:S2}) is denoted by $S_0$. The matrix
$S^{-1} e^{-i\lambda L}S e^{i\lambda L}$ can be represented as 
\begin{equation}
\label{eq:R2}
S^{-1} e^{-i\lambda L}S e^{i\lambda L}=e^{i  L V t} R_0 e^{-i L V t}
\end{equation}
where $R_0= S_0^{-1} e^{-i\lambda L}S_0e^{i\lambda L}$ is
time-independent. 
 It is
convenient to decompose $R_0$ into a product of upper and
lower-triangular matrices
\begin{equation}
\label{eq:R0-decomp}
R_0= \begin{pmatrix} 1& 0\\ \alpha & 1 \end{pmatrix} 
     \begin{pmatrix} a& 0\\ 0 & \frac1a \end{pmatrix} 
     \begin{pmatrix} 1& \beta\\ 0 & 1 \end{pmatrix}, 
\end{equation}
where
\begin{eqnarray}  
\label{eq:a}
a &=& |A|^2 e^{i\lambda} + |B|^2 \\
\alpha &=& \frac{A^{*}B(1-e^{i\lambda})}{a} \nonumber \\
\beta &=& \frac{-AB^{*}(1-e^{-i\lambda})}{a}. \nonumber
\end{eqnarray}

\subsection{Long time limit $t_f V \gg 1$}
The scattering matrix~(\ref{eq:S3}) is periodic with period $T=2\pi/V$
and in the limit $t_f V \gg 1$ we can apply the results of section~\ref{sec:periodic}.
The RH problem~(\ref{eq:RH-circle}) takes the form
\begin{equation}
  \label{eq:RH-biased}
  Y_-Y_+^{-1}=
    \begin{pmatrix} 1& 0\\ \frac{\alpha}{z} & 1 \end{pmatrix} 
     \begin{pmatrix} a& 0\\ 0 & \frac1a \end{pmatrix} 
     \begin{pmatrix} 1& \beta z\\ 0 & 1 \end{pmatrix}, 
\end{equation}
where $z=e^{i V t}$ and we used~(\ref{eq:R2}) and~(\ref{eq:R0-decomp}).

The solution of RH problem~(\ref{eq:RH-biased}) is obvious now
\begin{eqnarray*}
Y_-&=&\begin{pmatrix} 1&0\\ \frac{\alpha}{z} & 1 \end{pmatrix} \\
Y_+&=&\begin{pmatrix} 1& -\beta z\\ 0 & 1 \end{pmatrix} 
\begin{pmatrix} \frac1a & 0\\ 0 & a \end{pmatrix}.  
\end{eqnarray*}   
Substituting $Y_+$ into~(\ref{eq:chi-periodic}) we get
\begin{equation}
  \label{eq:contact-answer}
  \ln\chi(\lambda)=\frac{t_f V}{2\pi}\ln a 
\end{equation}
with $a$ given by~(\ref{eq:a}).  The above formula is LL's
answer\cite{LevitovJETP93} for the counting statistics of a biased
quantum point contact. Thus we have verified that measuring charge in one
lead after the leads are disconnected yields the same counting statistics
(in the limit $t_f V \gg 1$) as the measurement scheme based on a
coupling to a spin, as suggested by LL.

\section{Corrections to the Counting Statistics due to finite measuring  Time}
\label{correction}
\subsection{general considerations}
Eq~(\ref{eq:contact-answer}) gives the leading term in the expansion of
$\ln \chi$ in $t_f V$. The correction is proportional to $\ln t_f V$, and
to compute it we need to solve RH problem~(\ref{eq:RH-2})
\begin{equation}
  \label{eq:RH-switch}
  Y_- Y_+^{-1}= \begin{pmatrix} 1& 0\\ \alpha e^{- i V t} & 1 \end{pmatrix} 
     \begin{pmatrix} a& 0\\ 0 & \frac1a \end{pmatrix} 
     \begin{pmatrix} 1& \beta e^{i V t} \\ 0 & 1 \end{pmatrix} 
\end{equation}
This RH problem is no longer periodic, since we need to take into account
explicitly the finite length of the cut. We are not aware of an
analytical solution for RH problem~(\ref{eq:RH-switch}).  Instead, we
find an approximate solution in the limit $t_f V \gg 1$. To do this we
use the saddle-point method introduced for RH problems in
Ref.~\onlinecite{DeiftZhou}.  The idea is to modify RH
problem~(\ref{eq:RH-switch}) in such a way that the cuts become vertical
and the oscillating factors $e^{\pm i V t}$ turn into decaying ones.  To
implement the plan, we construct an auxiliary matrix-valued function $W$
and look for the solution in the form $Y=W\xi$. It turns out that $\xi$
obeys another RH problem, where all cuts are vertical, and the jump
functions decay along the cuts as $e^{-|t| V}$. As a result $\xi$ is
close to $1$ away from the branch points $0$ and $t_f$, and $W$ yields
the desired approximate solution to RH problem~(\ref{eq:RH-switch}).

\subsection{approximate solution of RH problem~(\ref{eq:RH-switch})}
\label{xi-approx}
It is convenient to introduce the function
\begin{equation}
  \label{eq:psi-solution}
  \psi(z) = \exp \left[  \frac{\ln a}{2\pi i} \left( \begin{smallmatrix}
      1 & 0\\
      0 & -1
    \end{smallmatrix} \right) \ln \frac{z}{z-t_f}  \right].
\end{equation}
which has the same region of analyticity as $Y$, tend to $1$ at infinity
and obeys
\begin{equation}
  \label{eq:RH-psi}
  \psi_- \psi_+^{-1} =      
\begin{pmatrix}
      a & 0\\
      0 & \frac1a
    \end{pmatrix}
\end{equation}
along the cut $(0,t_f)$.

\begin{figure}[htbp]
  \centering
  \includegraphics{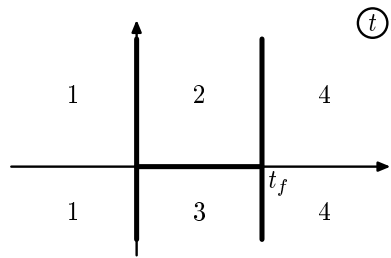}
  \caption[]{Regions of analyticity for $W(t)$.}
  \label{fig:W}
\end{figure}

The approximate solution $W(t)$ is constructed as analytic in the
complement of three cuts. The first two cuts are along the two vertical
lines $\re t=0$ and $\re t = t_f$ and the third cut is along the real
axis between $0$ and $t_f$ (see fig~\ref{fig:W}). The cuts divide the
complex plane into four regions : the half-plane $\re t < 0$ (region 1),
the upper part of the strip strip $ 0< \re t < t_f $ (region 2), the
lower part of the same strip (region 3) and the half-plane $\re t > t_f$
(region 4). Denoting the value of the function $W$ inside the region $i$
as $W_i(t)$ we define $W$ in the following way:
\begin{eqnarray*}
  W_1 &=& \psi, \quad  W_2 = 
 \begin{pmatrix}
     1 & \beta e^{i V t} \\ 
     0 & 1
\end{pmatrix}^{-1}\psi \\
W_4 &=& \psi, \quad 
W_3 = \begin{pmatrix}
 1                      & 0  \\
 \alpha e^{-i V t} & 1
\end{pmatrix} \psi. \\ 
\end{eqnarray*}
Comparing~(\ref{eq:RH-psi}) and~(\ref{eq:R0-decomp})
with~(\ref{eq:RH-switch}) we note the important property
\begin{equation}
  \label{eq:W-cut}
 W_-W_+^{-1}=Y_- Y_+^{-1}
\end{equation}
along the cut $(0,t_f)$.

We are now looking for the solution of the original RH
problem~(\ref{eq:RH-switch}) in the form $Y=W \xi$.
Combining~(\ref{eq:RH-switch}) with the definition of $W$ we conclude
that $\xi$ is analytic in each of the three regions $\re t<0$ (\ie region
1), $0< \re t < t_f$ (this is regions 2 and 3 combined together) and $\re
t > t_f$ (\ie region 4). Due to condition~(\ref{eq:W-cut}) the function
$\xi$ does not have a cut between regions 2 and 3. The other conditions
on the function $\xi$ come from the requirement that $Y$ is continuous on
the vertical cuts of $W$. Using the notation $\xi_i$ for the value of
$\xi$ in the region $i$, this condition gives on the upper half of the
cut $\re t = 0$
\begin{equation}
  \label{eq:RH-xi1}
  \xi_1 \xi_2^{-1} = W_1^{-1} W_2 = \psi^{-1}  \begin{pmatrix}
     1 & \beta e^{i V t} \\ 
     0 & 1
\end{pmatrix}^{-1} \psi,  
\end{equation}
and on the lower half of this cut
\begin{equation}
  \label{eq:RH-xi2}
  \xi_1 \xi_3^{-1} = W_1^{-1} W_3 = \psi^{-1}  \begin{pmatrix}
     1 & 0 \\
     \alpha e^{-i V t} & 1\\ 
\end{pmatrix} \psi. 
\end{equation}
Importantly, the right hand sides of
Eqs.~(\ref{eq:RH-xi1}),~(\ref{eq:RH-xi2}) are exponentially close to 1
when $t$ goes to infinity along the cuts. The difference is of the order
$e^{-|t|V}$ and can be neglected when $|t| \gg 1/V$. The same happens
along the other cut at $\re t = t_f$. Using this fact alone we prove (see
appendix~\ref{vertical-cuts}) that the function $\xi(t)$ is close to $1$
when $t$ is far away from $0$ and $t_f$. More precisely, in the region
\begin{equation}
  \label{eq:xi=1}
 |t| \gg \frac1V \hbox{ and } |t-t_f| \gg \frac1V
\end{equation}
$\xi$ is close to $1$, so $Y$ is close to $W$, as required.  Explicitly,
we obtain for $Y_+$
\begin{equation}
  \label{eq:Rplus}
Y_+(t) =  \left[ 
  \begin{array}[c]{rcl}
\psi_+(t) &\hbox{ when }& t< 0 \\
\begin{pmatrix}
     1 & -\beta e^{i V t} \\ 
     0 & 1
\end{pmatrix} \psi_+(t) &\hbox{  when }& 0 < t < t_f \\
\psi_+(t) &\hbox{ when }&  t_f < t
\end{array} \right.
\end{equation}
where $\psi$ is given by~(\ref{eq:psi-solution}). Eq.~(\ref{eq:Rplus}) is
not valid in the vicinity of the end points $0$ and $t_f$, where
conditions~(\ref{eq:xi=1}) are not met. It turns out that we do not need
the exact behavior of $Y_+$ in those regions to obtain the $\ln t_f V$
term in $\ln \chi $.

\subsection{Computation of the distribution function and
  interpretation of the result}
\label{two-sources}
The function $Y_+(t)$ enters $\ln \chi$ through the combination $\frac{d
  Y_+}{d t} Y_+^{-1}$ (see Eq.~\ref{eq:tr3}). There are, therefore, two
contribution to $\ln \chi$. The first contribution comes from
differentiating the factor $e^{i V t}$ in~(\ref{eq:Rplus}) and reproduces
Eq.~(\ref{eq:contact-answer}) which is linear in $t_f V$. The second
contribution comes from differentiating $\psi$ in~(\ref{eq:Rplus}) and is
proportional to $\ln t_f V$. Combining both contributions together we obtain
\begin{equation}
  \label{eq:chi-final}
  \ln \chi = \frac{t_f V}{2\pi} \ln a + \frac{\ln t_f V}{2 \pi^2} \ln^2 a,
\end{equation}
where $a$ is given by~(\ref{eq:a}). 

Finally, there are contributions from the regions $ t < 1/V$ and $ |t -
t_f| < 1/V$ where the asymptotic solution constructed above is not valid.
These contributions are of order $ \ln (\tau V)$, where $\tau$ is the
switching time. Assuming $\tau \sim 1/V$ these extra logarithms can be
neglected. Combining all the conditions specified above we find the
validity region for Eq.~(\ref{eq:chi-final})
\begin{equation}
  \label{eq:validity}
  \tau \sim 1/V \ll t_f. 
\end{equation}

The characteristic function~(\ref{eq:chi-final}) corresponds (with 
logarithmic accuracy) to the  probability distribution
\begin{equation}
  \label{eq:convolution-p-B}
  P(k)=\sum_{n}p(n)B(n,k),
\end{equation}
which is a convolution of a Gaussian  distribution for the number of attempts  
\begin{equation}
  \label{eq:p(N)-open}
  p(n)\propto e^{-\pi^2 (n-\frac{t_{f}V}{2\pi})^2/(4 \ln \frac{t_f V}{2\pi})}
\end{equation}
and the binomial distribution
\begin{equation}
  \label{eq:B(N,k)}
  B(n,k)=\frac{n!}{k! (n-k)!}
  |A|^{2k}|B|^{2(n-k)}.
\end{equation}
for the number of successes. The mean number of attempts is given by $t_f
V/2\pi$ and the variance is logarithmic $\sigma^2 \sim \ln \frac{t_f
  V}{2\pi} $.

There are, therefore, two sources of shot noise in a quantum point
contact. One source is the fluctuations in the number of incident
electrons (number of attempts to go through the contact) and the other is
the fluctuations in the number of reflections (number of failures).
According to (\ref{eq:convolution-p-B}) these two sources are
statistically independent. The physical picture of the two statistically
independent sources of quantum shot noise was suggested (but not proven)
by LL\cite{LevitovJETP93}.
Note, that in order to confirm the statistical independence we needed to
obtain the second term in Eq.~(\ref{eq:chi-final}) and analyze its
dependence on the transmission amplitude $A$.


\section{Distribution Function of the Dissipated Energy}
\label{sec:energy}
We mention without derivation that the characteristic function 
$\chi_e(\lambda)=\int P(E)e^{-i\lambda E} dE$
for the distribution of {\bf energy} dissipated during the pumping
\cite{adamov-muzykantskii}, is given by the expression
\begin{eqnarray}
  \label{eq:energy-answer}
  \ln \chi_e(\lambda) &=& \frac{i}{2\pi}\int_{0}^\lambda d\lambda 
  \int dt \tr\left\{ 
    \frac{dY_{+e}}{dt}Y_{+e}^{-1} R_e^{-1}\frac{dR_c}{d\lambda}
    \right\}\nonumber\\
    &-& i\lambda E_{av},
\end{eqnarray}
which is  very similar in form to Eq.~(\ref{eq:chi-general}). 
The analog of the Rieman-Hilbert problem (\ref{eq:RH-2}) is  
$Y_{-e}Y_{+e}^{-1}=R_e$, where
\begin{equation}
  \label{eq:R-e}
  R_e(\tau) = S^{-1}(\tau) S(\tau+\lambda).
\end{equation}
The average dissipated energy \cite{Makoshi:83} 
\begin{equation}
  \label{eq:E-av}
  E_{av} =   \frac{1}{4 \pi} \int_0^{t_f} \tr ( \frac{d S}{d \tau} \frac{d
    S^{-1}}{d \tau}) d \tau.
\end{equation}
appears analogously to  $Q_{av}$ during the regularisation of the determinant.

Similarly, the distribution function of the values of any single-particle
operator is given by expression similar to Eq.~(\ref{eq:chi-general}),~(\ref{eq:energy-answer}) with appropriate replacement of $R$ and $Q_{av}$.  

\section{conclusion}

It is clear from the derivation, that the charge operator $Q$ in
equation~(\ref{eq:chi-1}) can be replaced with any single-particle
operator. This will modify the counter-term~(\ref{eq:Q-av}), as well as
the relation~(\ref{eq:R}) between $R$ and $S$. Also
Eq.~(\ref{eq:chi-general} that relates the characteristic function with
the solution of the RH problem is modified. On the other hand, the
general form of the answer remains the same, \ie the distribution
function is still expressed as an integral over an expression containing
the solution of a particular matrix RH problem.

The general method is used to compute the distribution of the charge
transmitted through a biased quantum point contact in finite time.  We
confirm LL's picture of the two sources of the charge fluctuation (see
section~\ref{two-sources}) and prove that the transmission attempts are
statistically independent from the reflections when the total observation
time is much longer than inverse bias.

\appendix

\section{asymptotic expansion for~$\xi$}
\label{vertical-cuts}

To complete the discussion in section~\ref{xi-approx} we need to find the
function $\xi(t)$ with the following properties : 
\begin{itemize}
\item $\xi$ is analytic in the complement of the two vertical cuts $\re t =0 $ and
  $\re t = t_f$ 
\item $\xi$ obeys 
  \begin{equation}
 \label{eq:xi-appendix}
\xi_1 \xi_i^{-1} = W_1^{-1} W_i
\end{equation}
along the cut $\re t =0$, where $i=2,3$ and notations are the same as in
section~\ref{xi-approx}.
\item $\xi$ obeys $\xi_4 \xi_i^{-1}=W_4^{-1} W_i$ along the
  cut $\re t = t_f$.
\item $\xi \to 1 $ when $ |t| \to \infty $
\end{itemize}
In this appendix we show that $\xi$ is close to $1$ in the limit $V t_f
\gg 1$ everywhere except in the two small regions near $t=0$ and $t=t_f$

The general method can be illustrated by finding a function $\eta$ with
just one vertical cut along the upper part of imaginary axis ( $\re t =
0, \im t > 0 $) obeying the condition $ \eta_1 \eta_2^{-1} = W_1^{-1}
W_2$ along that cut. The case with two cuts is exactly the same.

Analyticity of $\eta$ together with the boundary condition at infinity
leads to the representation
$$
\eta(z) = 1+ \frac1{2\pi i} \int_{\re t' = 0; \im t' > 0 }
\frac{\eta(t'-0) - \eta(t'+0)}{t' - z} dt.
$$
Using the above we obtain the integral equation for the values of
$\eta_1(t)$ on the line $\re t =0$
\begin{eqnarray}
  \label{eq:RH-equation}
  && \eta_1 (t) = 1 +  \\
  && \frac1{2\pi i} \int_{\re t' = 0; \im t' > 0  }
  \frac{\left(1-W_2^{-1}(t')W_1(t')\right) \eta_1(t')}{t' - t + 0} d t'
  \nonumber
\end{eqnarray}
Because of exponential decay of the kernel along the cut
$$
1-W_2^{-1} W_1 = - e^{i V t} \psi^{-1}  \begin{pmatrix}
     0 & \beta  \\ 
     0 & 0
\end{pmatrix} \psi,
$$
the integral in Eq~(\ref{eq:RH-equation}) converges at $t' \sim 1/V
\ll t$, so in the limit $V t \gg 1$ we can neglect $t'$ in the
denominator and get the expansion
$$
\eta_1 = 1 + \frac{C}{Vt} + \cdots,
$$
where 
$$
C= \frac{V}{2\pi i} \int_{\re t' = 0; \im t' > 0 } \left(1-W_2^{-1}(t')W_1(t')\right)
\eta_1(t') d t' \sim 1
$$
is some constant.

\end{document}